\documentclass[aps,prb,twocolumn,showpacs,floatfix]{revtex4}

\usepackage{amsmath}
\usepackage{amsfonts}
\usepackage{amssymb}
\usepackage{graphicx}
\usepackage{bm}

\begin{document}

\title{Greenberger-Horne-Zeilinger States in Quantum Dot Molecule}

\author{Anand Sharma}
\email{anand.sharma@nrc-cnrc.gc.ca}
\altaffiliation{Present address : Department of Physics, 82 University Place, University of Vermont, Burlington, USA.}

\author{Pawel Hawrylak}
\email{pawel.hawrylak@nrc-cnrc.gc.ca}
\affiliation{Department of Physics, University of Ottawa, Ottawa, Canada, K1N6N5}
\affiliation{Institute for Microstructural Sciences, National Research Council of Canada, Ottawa, Canada, K1A0R6}

\begin{abstract}
We present a microscopic theory of a lateral quantum dot molecule in a radial magnetic field with a Greenberger- Horne- Zeilinger (GHZ) maximally entangled three particle groundstate. The quantum dot molecule consists of three quantum dots  with one electron spin each forming a central equilateral triangle. The anti-ferromagnetic spin-spin interaction is changed to the ferromagnetic interaction by additional doubly occupied quantum dots, one dot near each side of a triangle. The magnetic field is provided by micro-magnets. The interaction among the electrons is described within an extended Hubbard Hamiltonian and electronic states are obtained using configuration interaction approach. The set of parameters is established for which the ground state of the molecule in a radial magnetic field is well approximated by a GHZ state.
\end{abstract}

\date{\today}


\maketitle

\section{\label{sec:intro} Introduction}

\indent In a recent work Roethlisberger et al., Ref.~\onlinecite{roethlisberger}, proposed a scheme to generate maximally entangled Greenberger- Horne- Zeilinger~\cite{ghz} (GHZ) state as a ground state of a three spin system. The spins were arranged in a triangular geometry and assumed to interact ferromagnetically. With ferromagnetic interaction the degenerate ground state is maximally spin polarized, with two $ |\uparrow \uparrow \uparrow>$ and $|\downarrow \downarrow \downarrow>$ configurations. When radial in- plane magnetic field is applied the two spin polarized configurations form the two GHZ states $ |GHZ^{\pm}>={1 \over{\sqrt{2}}}(|\uparrow \uparrow \uparrow> \pm |\downarrow \downarrow \downarrow>)$. Such GHZ ground states of spin molecules could be used as long lived sources of entanglement~\cite{bohm,nielsenchuang}. One can envisage using lateral triple quantum dot molecules\cite{gaudreau,korkusinski1,gimenez} with one electron on each dot to realize a three spin system. However, the spin-spin interaction in quantum dot molecules is necessarily anti-ferromagnetic and such a simple quantum dot molecule is not possible. Here we propose more complex molecule which uses auxiliary quantum dots to effectively change the spin-spin interaction to ferromagnetic and generate the maximally entangled GHZ state as its ground state.\\

\indent Theoretical proposals to realize GHZ states in solid state systems include, e.g., spin systems~\cite{roethlisberger}, excitons in coupled dots~\cite{quiroga}, two-level atoms in a non-resonant cavity~\cite{zheng} and superconducting flux qubits~\cite{kim}. The GHZ states have been experimentally realized using photons~\cite{bouwmeesterpan}, in atomic systems using three Rydberg atoms~\cite{rauschenbeutel} and very recently maximally entangled GHZ states have been realized in solid state
system using superconducting qubits~\cite{neeley,dicarlo}. We refer the reader to Refs.~\onlinecite{neeley,dicarlo} for full initialization, characterization and read out of GHZ states.\\

\indent In this paper, we propose a theoretical model of six coupled quantum dots~\cite{korkusinski1,gimenez,hawrylak1,jacaksachrajda} with interacting electrons to generate the GHZ state as a ground state. The model consists of three equally spaced quantum dots with one electron spin each. The anti-ferromagnetic interaction of each pair of localized spins is modified by connecting each pair of spins to an auxiliary quantum dot with two electrons. In Ref.~\onlinecite{shim2} it was shown that the effective interaction between two electrons in two quantum dots i.e., one electron in each of the two quantum dots, can be changed from anti- ferromagnetic to ferromagnetic by connecting them to a third (auxiliary) doubly occupied quantum dot and applying a bias to the auxiliary quantum dot. Thus it was shown that there exists a set of parameters for which such a triple quantum dot molecule with four electrons can be thought of as effectively two localized electron spins interacting ferromagnetically. Following this idea we propose here a quantum dot molecule consisting of three dots with one electron each and three auxiliary dots with two electrons each which effectively realizes the ferromagnetically coupled three spin cluster. Using Hubbard model and exact diagonalization techniques we show that the ground state of such a molecule in a radial magnetic field is indeed a GHZ state.\\

\indent The structure of the paper is organized as follows. In Section~\ref{sec:tpsys} we examine the proposed setup for generating highly entangled states in a three spin system interacting ferromagnetically with each other and coupled to a radial in- plane magnetic field. We clarify the role of applied magnetic field in generating the GHZ states using the degenerate perturbation theory. In Section~\ref{sec:qdm} we propose a quantum dot molecule consisting of six quantum dots with nine electrons and micro-magnets to generate radial in- plane magnetic field creating the maximally entangled GHZ state as its ground state. We describe the system, its Hamiltonian and ground state phase diagram. The last section~\ref{sec:sumconc} contains summary and conclusions.\\

\section{\label{sec:tpsys} Generating GHZ state in a three spin system}
\begin{figure}[!htbp]
\centering
\vspace*{0.6cm}
\includegraphics[height=5.0cm,width=8.5cm]{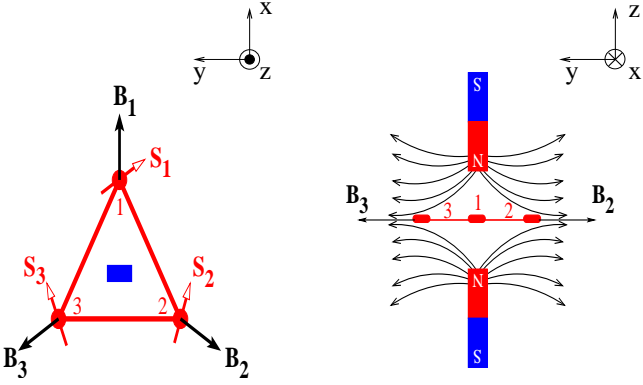}
\vspace*{0.4cm}
\caption{\label{fig:tmod} (Color online) Schematic showing two views of three localized electron spins \textbf{S}$_{\textrm{i}}$ (i=1,2,3) arranged in a triangular geometry and interacting ferromagnetically with each other. The two bar magnets at the triangle centroid are placed on top of each other generating a radial magnetic field \textbf{B}$_{\textrm{i}}$. Figure on the left shows the top view of such a system and the right panel exhibits the lateral view. The magnetic field lines are also shown.}
\end{figure}

We begin with an analysis of the three spin system to create the maximally entangled states as outlined in Ref.~\onlinecite{roethlisberger}. The three spins $\frac{1}{2}$ placed at the corners of an equilateral triangle lying in the \textit{xy} plane as shown in left panel of Fig.~\ref{fig:tmod} are described by an isotropic Heisenberg Hamiltonian,

\begin{equation}\label{eq:isoheisham}
\textrm{H}_{\textrm{H}} = -\textrm{J} \sum_{\textrm{i=1}}^{3} \textbf{S}_{\textrm{i}} \cdot \textbf{S}_{\textrm{i+1}} \hspace*{1.35cm} (\textbf{S}_{4} = \textbf{S}_{1})
\end{equation}

where it is assumed that the exchange coupling $\textrm{J}$ is ferromagnetic ($\textrm{J}>0$) and $\textbf{S}_i = {\hbar \over{2}}(\sigma^{\textrm{x}}_{\textrm{i}},\sigma^{\textrm{y}}_{\textrm{i}}, \sigma^{\textrm{z}}_{\textrm{i}})$ is a vector consisting of Pauli matrices,
$\sigma^{\textrm{x}}_{\textrm{i}} =
\left( \begin{array}{cc}
0 & 1 \\
1 & 0 \\
\end{array} \right)$,
$\sigma^{\textrm{y}}_{\textrm{i}} =
\left( \begin{array}{cc}
0 & -i \\
i & 0 \\
\end{array} \right)$ and
$\sigma^{\textrm{z}}_{\textrm{i}} =
\left( \begin{array}{cc}
1 & 0 \\
0 & -1 \\
\end{array} \right)$
acting on sites i (=1,2,3). Here $\hbar$ is reduced Planck constant.

\indent We write the Hamiltonian, given in Eq.~\ref{eq:isoheisham}, in the basis of tensor product of states at each site. For example, the Pauli matrices are written in such a local basis. Thus the Hamiltonian in the basis set $\big\{ |\uparrow \uparrow \uparrow>, |\uparrow \uparrow \downarrow>, |\uparrow \downarrow \uparrow>, |\uparrow \downarrow \downarrow>, |\downarrow \uparrow \uparrow>, |\downarrow \uparrow \downarrow>, |\downarrow \downarrow \uparrow> \textrm{and} |\downarrow \downarrow \downarrow>\big\}$ is a 8 x 8 matrix.\\

\indent Exact diagonalization of the Hamiltonian  matrix in the basis of $8$ states yields two quadruplet levels corresponding to total spin $S=3/2$ and total spin $S=1/2$. We write the eigenstates of the Hamiltonian in Eq.~\ref{eq:isoheisham} as $|\alpha, \textrm{S}, \textrm{S}^{\textrm{z}}_{\textrm{t}}>$ where $\alpha$ is the index of the eigenstate, $\textrm{S}$ the total spin and $\textrm{S}^{\textrm{z}}_{\textrm{t}}$ the z-component of total spin of the system. \\

\indent The ground state $S=3/2$ quadruplet consists of the following four eigenstates with eigenvalue $-\frac{3}{4}\textrm{J}$,

\begin{equation}\label{eq:bas1}
\bigg|1, \frac{3}{2}, \frac{3}{2}\bigg> = |\uparrow\uparrow\uparrow> 
\end{equation}
\begin{equation}\label{eq:bas2}
\bigg|2, \frac{3}{2}, -\frac{3}{2}\bigg> = |\downarrow\downarrow\downarrow>
\end{equation}
\begin{equation}\label{eq:bas3}
\bigg|3, \frac{3}{2}, \frac{1}{2}\bigg> = \frac{1}{\sqrt{3}}\big[ |\downarrow\uparrow\uparrow> + |\uparrow\downarrow\uparrow> + |\uparrow\uparrow\downarrow>\big]
\end{equation}
\begin{equation}\label{eq:bas4}
\bigg|4, \frac{3}{2}, -\frac{1}{2}\bigg> = \frac{1}{\sqrt{3}}\big[ |\uparrow\downarrow\downarrow> + |\downarrow\uparrow\downarrow> + |\downarrow\downarrow\uparrow>\big]
\end{equation}

The four states correspond to the two distinct classes~\cite{dueracin} of highly entangled states of a tripartite qubit system. As will be shown below, the mixed $|$1$>$ and $|$2$>$ states form the two states of the Greenberger-Horne-Zeilinger class $\big[|\textrm{GHZ}^{\pm}>=\frac{1}{\sqrt{2}}\big(|\uparrow\uparrow\uparrow> \pm |\downarrow\downarrow\downarrow>\big)\big]$ while $|$3$>$ and $|$4$>$ belong to the W class~\cite{neeley}.\\

\indent The excited quadruplet with total spin $S=1/2$ and eigenvalue $+\frac{3}{4}\textrm{J}$ comprises of the following four chiral states~\cite{hawrylak1,subrahmanyam},

\begin{equation}\label{eq:bas5}
\bigg|5, \frac{1}{2}, \frac{1}{2}\bigg> = \frac{1}{\sqrt{3}}\big[ |\downarrow\uparrow\uparrow> +  e^{\frac{i2\pi}{3}}|\uparrow\downarrow\uparrow> + e^{\frac{i4\pi}{3}}|\uparrow\uparrow\downarrow>\big]
\end{equation}
\begin{equation}\label{eq:bas6}
\bigg|6, \frac{1}{2}, \frac{1}{2}\bigg> = \frac{1}{\sqrt{3}}\big[ |\downarrow\uparrow\uparrow> +  e^{-\frac{i2\pi}{3}}|\uparrow\downarrow\uparrow> + e^{-\frac{i4\pi}{3}}|\uparrow\uparrow\downarrow>\big]
\end{equation}
\begin{equation}\label{eq:bas7}
\bigg|7, \frac{1}{2}, -\frac{1}{2}\bigg> = \frac{1}{\sqrt{3}}\big[ |\uparrow\downarrow\downarrow> + e^{\frac{i2\pi}{3}}|\downarrow\uparrow\downarrow> + e^{\frac{i4\pi}{3}}|\downarrow\downarrow\uparrow>\big]
\end{equation}
\begin{equation}\label{eq:bas8}
\bigg|8, \frac{1}{2}, -\frac{1}{2}\bigg> = \frac{1}{\sqrt{3}}\big[ |\uparrow\downarrow\downarrow> + e^{-\frac{i2\pi}{3}}|\downarrow\uparrow\downarrow> + e^{-\frac{i4\pi}{3}}|\downarrow\downarrow\uparrow>\big]
\end{equation}

\indent These chiral states are the eigenstates of chirality operator $\chi=\textbf{S}_{1} \cdot (\textbf{S}_{2} \times \textbf{S}_{3})$ as described in Ref.~\onlinecite{hsieh}. The states have the spin ( $\uparrow$ and $\downarrow$ ) currents going in two different ( $\circlearrowright$ and $\circlearrowleft$ ) directions. \\

\indent For simplicity we shall shift the overall energy scale by $\frac{3}{4}\textrm{J}$ with the energy of the ground state quadruplet as the reference energy. In order to obtain one of the two GHZ states as a ground state of the system we need to apply a perturbation which will mix the states $|$1$>$ and $|$2$>$ while separating them from the remaining two W states, $|$3$>$ and $|$4$>$, thus splitting the degenerate eigenspace of the lower quadruplet. As was shown in Ref.~\onlinecite{roethlisberger} this can be accomplished by  applying an in- plane radial magnetic field. \\

\indent We now apply the radial magnetic field $\textbf{B}$ as shown in Fig.~\ref{fig:tmod}. The Hamiltonian $\textrm{H}_{\textrm{B}} = \textrm{g}_{\textrm{e}} \mu_{\textrm{B}} \sum_{i} \textbf{B}_{\textrm{i}} \cdot \textbf{S}_{\textrm{i}}$ describing the coupling of the spins to an externally applied in- plane radial magnetic field generated using the two bar magnets placed with similar poles~\cite{note1} facing each other as shown in the right panel of Fig.~\ref{fig:tmod}, reads

\begin{equation}\label{eq:radham}
\textrm{H}_{\textrm{B}} = \textrm{b} \bigg[ \sigma^{\textrm{x}}_{1} \bigg] + \textrm{b} \bigg[ -\frac{1}{2}\sigma^{\textrm{x}}_{2} -\frac{\sqrt{3}}{2}\sigma^{\textrm{y}}_{2} \bigg] + \textrm{b} \bigg[ -\frac{1}{2}\sigma^{\textrm{x}}_{3} +\frac{\sqrt{3}}{2} \sigma^{\textrm{y}}_{3} \bigg]
\end{equation}

where $\textrm{b} = \frac{\textrm{g}_\textrm{e}\mu_{\textrm{B}}\textrm{B}\hbar}{2}$ is an effective magnetic field with $\textrm{g}_{\textrm{e}}$ an effective electron g- factor in the plane of a quantum dot, $\mu_{\textrm{B}}$ the Bohr magneton and B the strength of the radial magnetic field at the quantum dot position. The effective magnetic field 'b' is simply related to Zeeman energy splitting $2 \textrm{b} = \Delta \textrm{E}_\textrm{z}$ for a given external magnetic field 'B'. And $\textbf{S}_{\textrm{i}}$ is a vector denoting the spin at quantum dot ``i'' with Pauli matrices as components along x and y directions.\\

\indent The complete Hamiltonian of the three spin system in an external magnetic field is given as

\begin{equation}\label{eq:fullham}
\textrm{H} = \textrm{H}_{\textrm{H}} + \textrm{H}_{\textrm{B}}.
\end{equation}

\indent The physical meaning of the radial magnetic field is best described by its action on the spin polarized state:

\begin{equation}\label{eq:1to6}
\textrm{H}_{\textrm{B}} |\uparrow\uparrow\uparrow> =  \frac{1}{\sqrt{3}}\big[ |\downarrow\uparrow\uparrow> +  e^{-\frac{i2\pi}{3}}|\uparrow\downarrow\uparrow> + e^{-\frac{i4\pi}{3}}|\uparrow\uparrow\downarrow>\big]
\end{equation}

\indent Thus the radial in- plane magnetic field simultaneously flips the spins and adds a phase factor in such a way as to generate a chiral state $|6>$ with $\textrm{S}=1/2, \textrm{S}^{\textrm{z}}_{\textrm{t}} = +1/2$ and momentum $k = 2 \pi/3$. In a similar way $\textrm{H}_\textrm{B}$ couples state $|6>$ ($\textrm{S}^{\textrm{z}}_{\textrm{t}} = +1/2, k = 2\pi/3$) with state $|7>$ ( $\textrm{S}^{\textrm{z}}_{\textrm{t}} = -1/2, k = -2 \pi/3$ ) and state $|7>$ with spin polarized state $|2>$ with ($\textrm{S}^{\textrm{z}}_{\textrm{t}} = -3/2, k = 0$). The transition from state $|1>$ to state $|2>$ is a third order process in $b$. The final effective Hamiltonian for the two spin polarized states ($|$1$>$ and $|$2$>$) and the two chirality states ($|$6$>$ and $|$7$>$) reads,
\begin{center}
$\left(\begin{array}{cc|cc}
0 & 0 & \textrm{b}\sqrt{3} & 0 \\
0 & 0 & 0 & \textrm{b}\sqrt{3} \\
\hline
\textrm{b}\sqrt{3} & 0 & \frac{3}{2}\textrm{J} & \textrm{2b} \\
0 & \textrm{b}\sqrt{3} & \textrm{2b} & \frac{3}{2}\textrm{J} \\
\end{array}\right)$\\
\end{center}

\indent With the radial magnetic field (second term in Eq.~\ref{eq:fullham}) treated as a perturbation, the degenerate perturbation theory~\cite{loewdin} leads to an effective 2 x 2 Hamiltonian matrix ($\mathnormal{H}^{eff}$) in the subspace of spin polarized S=$\frac{3}{2}$ states,

\begin{equation}\label{eq:effham}
\mathnormal{H}^{eff} =
\left( \begin{array}{cc}
0 & \frac{8\textrm{b}^{3}}{3\textrm{J}^{2}} \\
\frac{8\textrm{b}^{3}}{3\textrm{J}^{2}} & 0 \\
\end{array} \right)
\end{equation}

\indent The effective Hamiltonian is that of a two level system with tunneling $T= \frac{8\textrm{b}^{3}}{3\textrm{J}^{2}}$ proportional to the third power of the external magnetic field 'b' . After diagonalization of the matrix in Eq.~\ref{eq:effham} we obtain as the ground state the GHZ state $ |GHZ^{+}>={1 \over{\sqrt{2}}}(|\uparrow \uparrow \uparrow> + |\downarrow \downarrow \downarrow>$ with energy $-\frac{8\textrm{b}^{3}}{3\textrm{J}^{2}}$, separated from the second GHZ state by $2 T = \frac{16\textrm{b}^{3}}{3\textrm{J}^{2}}$ as also obtained in Ref.~\onlinecite{roethlisberger}.\\

\indent The energy splitting of the two GHZ states can be also written in terms of exchange coupling $\textrm{J}$ and the Zeeman energy as $ 2 T = \frac{2}{3} (\Delta \textrm{E}_\textrm{z})(\frac{\Delta \textrm{E}_\textrm{z}}{\textrm{J}})^2$. For GaAs $\Delta \textrm{E}_\textrm{z}$=20 $\mu$eV in the field of B=1 Tesla. The exchange coupling measured for coupled lateral GaAs quantum dots is comparable to Zeeman splitting as J$\sim$20 $\mu$eV~\cite{johnson}. Hence in lateral devices the two GHZ states can be separated on the order of Zeeman energy, i.e., several $\mu$ eV. The full discussion of physical parameters in gated and self-assembled quantum dot is deferred to the end of Section~\ref{sec:qdm}.\\

\indent In the next section we discuss how spins of electrons in a quantum dot molecule can be used to realize the GHZ states in solid state quantum dot molecules. \\

\section{\label{sec:qdm} Quantum Dot Molecule}

\indent A lateral triple quantum dot molecule (TQDM)~\cite{korkusinski1,gimenez,shim2,gaudreau,schroeer,hawrylak2,delgado} with one electron per dot is the simplest realization of the three spin system  discussed in the previous section. Assuming a single orbital per dot and arbitrary occupation, as shown in Refs.~\onlinecite{korkusinski1,gimenez} the quantum dot molecule can be described by an extended Hubbard Hamiltonian,

\begin{align}\label{eq:tqdham}
\mathcal{H}^{\textrm{QDM}} \nonumber
& = \sum_{\substack{\textrm{i=1} \\ \tau}}^{\textrm{N}} \epsilon_{\textrm{i}} \textrm{n}_{\textrm{i}\tau} - \sum_{\substack{\textrm{i,j=1} \\ \tau}}^{\textrm{N}} \textrm{t}_{\textrm{ij}} ( \textrm{c}_{\textrm{i}\tau}^{\dagger} \textrm{c}_{\textrm{j}\tau} + h.c.) + \textrm{U} \sum_{\substack{\textrm{i=1} \\ \tau}}^{\textrm{N}} \textrm{n}_{\textrm{i}\tau} \textrm{n}_{\textrm{i}-\tau} \nonumber \\
& + \sum_{\textrm{i,j=1}}^{\textrm{N}} \textrm{V}_{\textrm{ij}} \rho_{\textrm{i}} \rho_{\textrm{j}}
\end{align}

where N is the number quantum dots ( N$=3$ for TQDM), c$_{\textrm{i}\tau}^{\dagger}$ (c$_{\textrm{i}\tau}$) is the creation (annihilation) operator of an electron with spin $\tau (= \uparrow,\downarrow)$ at i-th QD , n$_{\textrm{i}\tau}$(= c$_{\textrm{i}\tau}^{\dagger}$c$_{\textrm{i}\tau}$) is the spin- dependent electron occupation number, $\rho_{\textrm{i}}$(=$\sum_{\tau} \textrm{n}_{\textrm{i}\tau}$) is the total occupation number at site ``i'', $\epsilon_{\textrm{i}}$ is the on- site energy, t$_\textrm{ij}$ is the inter- dot tunneling matrix element between nearest neighboring dots ``i'' and ``j'', U$_{\textrm{i}}$ is the intra- dot and V$_{\textrm{ij}}$ the inter- dot Coulomb repulsion between the nearest neighboring dots and \textit{h.c.} means hermitian conjugate. \\

\indent At half-filling the  Hubbard Hamiltonian reduces to an isotropic Heisenberg Hamiltonian but with anti-ferromagnetic exchange coupling~\cite{korkusinski1,gimenez}, $\sim 4 \textrm{t}^2 /\textrm{U}$, and the ground state (GS) is in the subspace of total spin S=$\frac{1}{2}$. Thus it is not possible to tune the GS of the system to a maximally spin polarized subspace of S=$\frac{3}{2}$ where the GHZ states reside. However, it was shown~\cite{korkusinski1,shim2} that triple quantum dot molecule with four electrons, which corresponds to two holes, has a spin polarized ground state. When one of the dots is biased, it contains two spin singlet electrons while the two remaining electrons, localized on the two remaining dots, are coupled ferromagnetically. Biasing the dot further effectively decouples this dot and leaves the remaining two dots with one electron each in a spin singlet state, i.e., coupled anti-ferromagnetically.\\

\indent Hence it is in principle possible to design a half- filled TQDM combined with auxiliary doubly occupied quantum dots in order to generate an effective ferromagnetic coupling of spins in the TQDM. If this is accomplished then we can realize the GHZ states in such a quantum dot molecule by applying a radial in- plane magnetic field as explained in the previous section. \\

\begin{figure}[!htbp]
\centering
\vspace*{0.6cm}
\includegraphics[height=5.0cm,width=5.0cm]{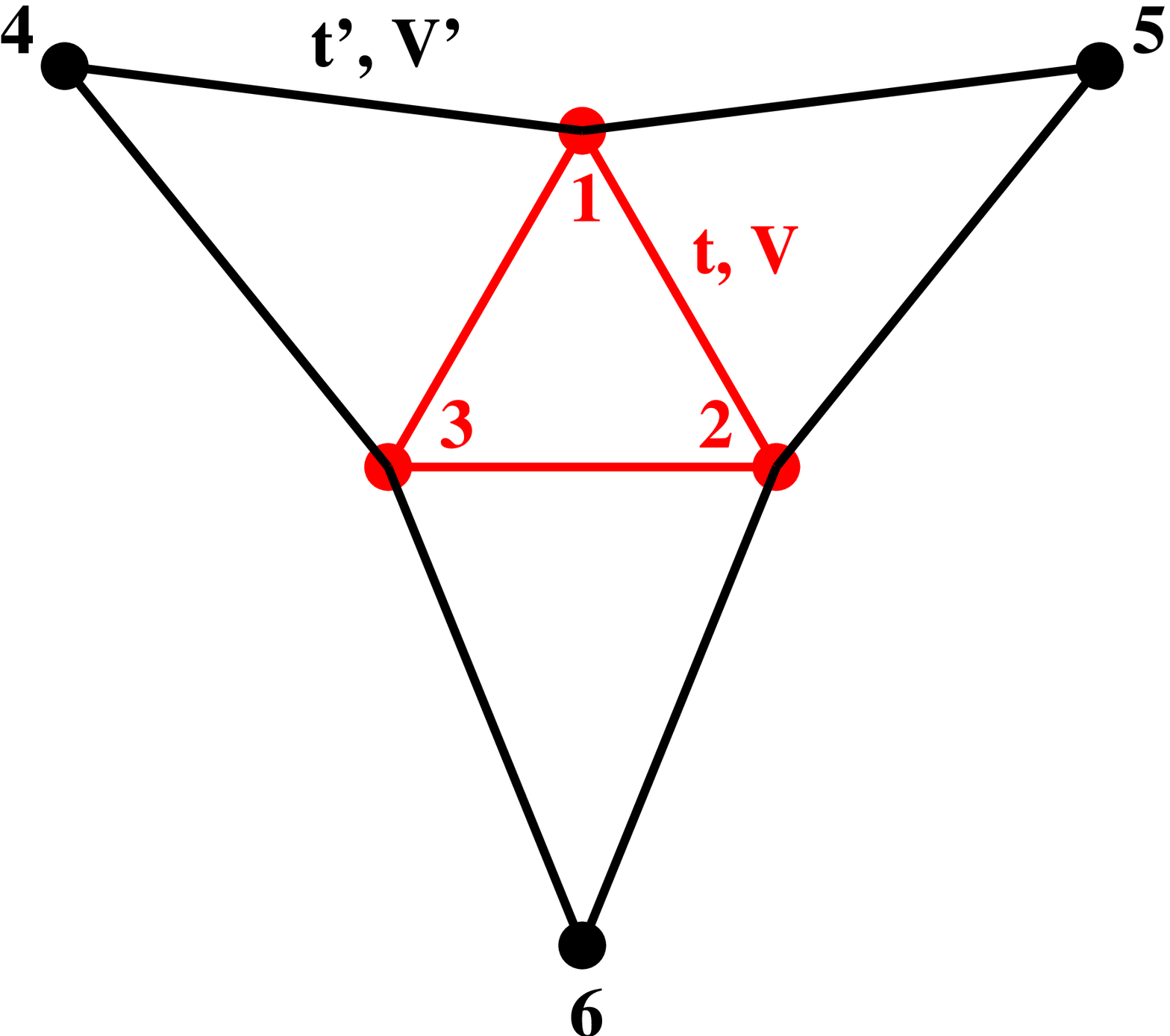}
\hfill
\includegraphics[height=2.75cm,width=2.75cm]{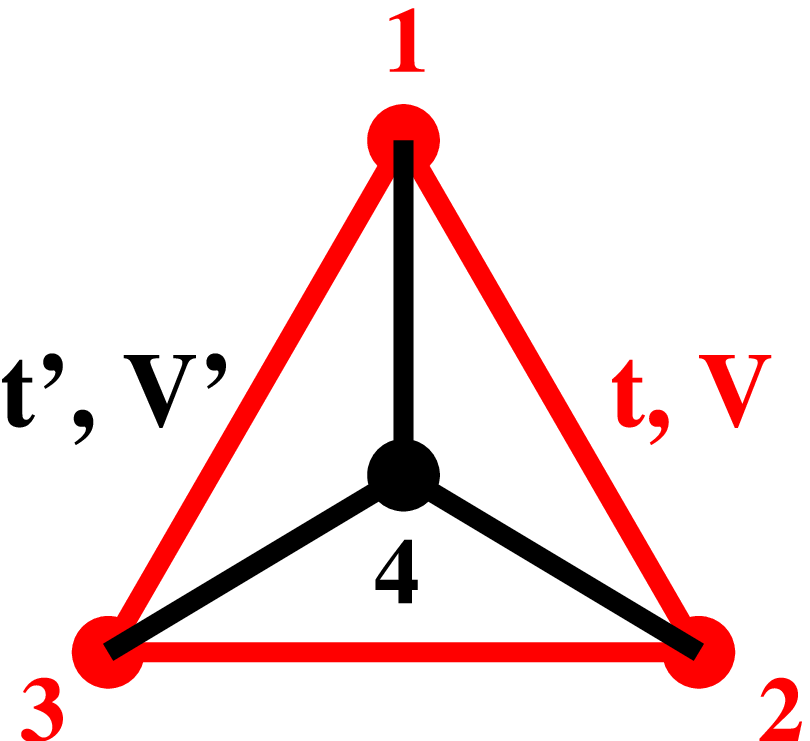}
\vspace*{0.4cm}
\caption{\label{fig:sfqdm} (Color online) \textit{Left}: Schematic of a six dot molecule with dots labelled (1,2,3) forming the central triple quantum dot molecule and dots (4,5,6) are the auxiliary dots. \textit{Right}: A four dot molecule structure with dot 4 being an additional dot introduced to the triple quantum dot molecule (1,2,3).}
\end{figure}

\indent In what follows we propose two design structures, namely a six and a four quantum dot molecule as shown on the left and right side of Fig.~\ref{fig:sfqdm} respectively. While studied both, we discuss in detail the six quantum dot molecule because it might be difficult to fabricate a triangular quantum dot structure with a central auxiliary dot in the middle.\\

\indent We form a six quantum dot molecule (SQDM) by bringing three doubly occupied dots labelled (4,5,6) close to the  central~\cite{note2} triangle (1,2,3) of singly occupied dots, for a total of nine electrons. The Hamiltonian for SQDM, $\mathcal{H}^{\textrm{SQD}}$, is given in Eq.~\ref{eq:tqdham} for N=6. We consider t$_{\textrm{ij}}$=t and V$_{\textrm{ij}}$=V for central triangle \{i,j\} = \{1,2,3\} and t$_{\textrm{ij}}$=t' and V$_{\textrm{ij}}$=V' for all other dots as shown in left panel of Fig.~\ref{fig:sfqdm}.\\

\indent We create all $220$ $\textrm{N}_\textrm{e}=9$ electron configurations $\prod_{\textrm{i},\sigma} \textrm{c}_{\textrm{i},\sigma}^+ |0>$, construct the Hamiltonian matrix and diagonalize it numerically to obtain the energy spectrum and eigenvectors. In the absence of the magnetic field  the z- component of the total spin is a good quantum number giving the Hamiltonian matrix in a block diagonal form. The Hilbert space dimensions of the blocks in the subspace of z- component of total spin S$_{\textrm{t}}^{\textrm{z}}$=$\pm \frac{1}{2}$  and S$_{\textrm{t}}^{\textrm{z}}$=$\pm \frac{3}{2}$ are 90 and 20 respectively. Each block is numerically diagonalized so as to obtain the GS of the system as a function of the Hubbard parameters. \\

\indent In our calculations we assume $\epsilon_{1} = \epsilon_{2} = \epsilon_{3} = 0$ i.e., the dots $(1,2,3)$ are on resonance. The energy $\epsilon_{4} = \epsilon_{5} = \epsilon_{6} = \epsilon$ of auxiliary dots is varied by applied gate voltage. We consider the regime U $\gg$ t,V with constant value of U=2.0, V=0.1 as considered earlier~\cite{shim1}. We vary inter- dot Coulomb repulsion and tunneling between central and auxiliary dots as 0 $<$ V' $<$ V and 0 $<$ t' $<$ t. All the parameters are in the unit of effective Rydberg defined by $\textit{Ry} = m^{*}e^{4} / 2\epsilon^{2}\hbar^{2}$ where $m^{*}$ is the electron effective mass, $e$ the electron charge, $\epsilon$ the dielectric constant of a material. For example in case of GaAs the effective Rydberg is estimated to be about 6 meV.\\

\begin{figure}[htbp]
\centering
\vspace*{0.6cm}
\includegraphics[height=6.0cm,width=7.0cm]{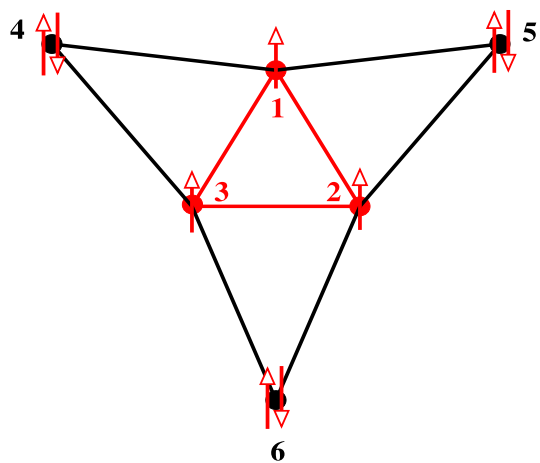}
\vspace*{0.4cm}
\caption{\label{fig:ccfsdm} (Color online) Dominant charge configuration in the subspace of total spin S=$\frac{3}{2}$ and z- component of total spin S$_{\textrm{t}}^{\textrm{z}}$=$\frac{3}{2}$ for a six dot molecule with nine electrons.}
\end{figure}

\indent The GS of the system is a linear superposition of the basis used to construct the Hamiltonian matrix in that given subspace. For certain parameters the GS contains the spin polarized configuration with spins localized on the three dots. This is illustrated in Fig.~\ref{fig:ccfsdm} which shows the dominant configuration, with probability 0.9289 for the GS in the subspace of total spin S=$\frac{3}{2}$ and z- component of total spin S$_{\textrm{t}}^{\textrm{z}}$=$\frac{3}{2}$ obtained for parameters $\epsilon = -0.05$, U=2.0, V=0.1, t=0.05, V'=0.094, t'=0.05. It is clearly seen that the three electron spins are localized on the triangle of dots $(1,2,3)$ forming the ferromagnetically coupled three spin system and the auxiliary dots $(4,5,6)$ are doubly occupied, driving the GS of the quantum dot molecule to be in the maximally spin polarized subspace. \\

\begin{figure*}[!htbp]
\includegraphics[height=11.0cm,width=14.0cm]{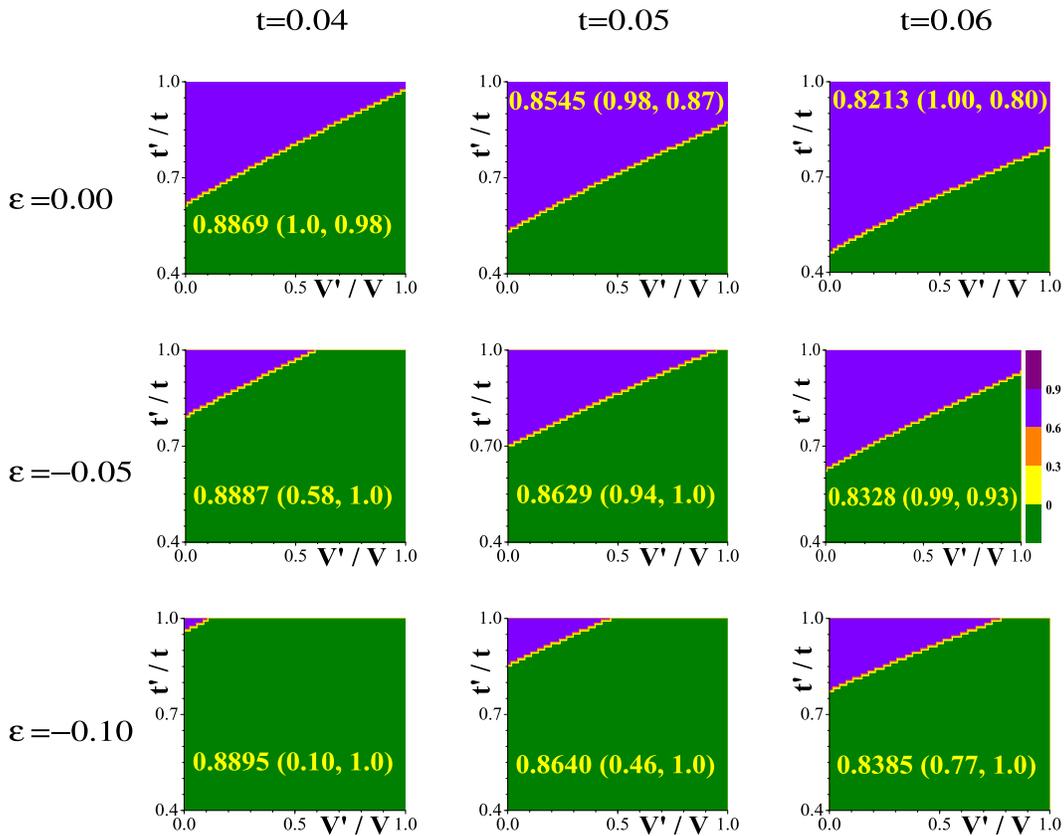}
\caption{\label{fig:phdia} (Color online) Phase diagram of ground state (GS) overlap probability with a state of dominant charge configuration in the subspace of S=$\frac{3}{2}$ for a six quantum dot molecule is shown as a function of t'/t and V'/V. The individual view graphs are for different values of hopping element, t and the on- site energy, $\epsilon$, at dots (4,5,6). The lower area in each graph (green in color) denotes the GS is in subspace of total spin S=$\frac{1}{2}$ and its overlap is zero. The number in each graph corresponds to the maximum value of the overlap probability at the point given in the bracket.}
\end{figure*}

\indent We define a state corresponding to the dominant charge configuration in the subspace of z- component of total spin S$_{\textrm{t}}^{\textrm{z}}$=$\frac{3}{2}$ as,

\begin{equation}\label{eq:dcfu}
|\textrm{US}> = | (\uparrow_{1}\uparrow_{2}\uparrow_{3})(\uparrow_{4}\downarrow_{4}\uparrow_{5}\downarrow_{5}\uparrow_{6}\downarrow_{6})>,
\end{equation}

with similar definition for the dominant charge configuration in the subspace of z- component of total spin S$_{\textrm{t}}^{\textrm{z}}$=$-\frac{3}{2}$ as,

\begin{equation}\label{eq:dcfd}
|\textrm{DS}> = | (\downarrow_{1}\downarrow_{2}\downarrow_{3})(\uparrow_{4}\downarrow_{4}\uparrow_{5}\downarrow_{5}\uparrow_{6}\downarrow_{6})>,
\end{equation}

\indent We now determine the set of parameters for which the SQDM describes the three spin system well, i.e. we calculate the overlap probability, at zero magnetic field ($\textrm{B}=0$),

\begin{equation}\label{eq:ovpwtbf}
\textrm{P}_{\textrm{B=0}} = |<\textrm{GS}|\textrm{US}>|^{2}
\end{equation}

of the numerically obtained GS with state $|\textrm{US}\rangle$ as a function of parameters. The non-zero overlap probability  as a function of parameters t,t',V' and $\epsilon$ for fixed U,V, determines the phase diagram, region in parameter space where the three electrons are localized on the three dots and interact ferromagnetically. The result of numerical calculation for U=2.0 and V=0.1 is shown in Fig.~\ref{fig:phdia}.\\

\indent We observe that for a fixed value of t, if we vary $\epsilon$ the region of GS in maximally spin polarized subspace gets reduced but the maximum value of the overlap probability increases marginally. This happens because as we lower the on- site energies of the auxiliary dots, the three central dots become isolated from the auxiliary system and behave like half- filled TQDM. And so a higher value of t' is required to have the ground state in maximally spin polarized subspace. Hence the region belonging to the total spin S=$\frac{3}{2}$ decreases.\\

\indent It is also found that for a fixed $\epsilon$, varying $\textrm{t}$ decreases the maximum value of overlap probability but the region of GS in maximally spin polarized subspace increases. It is because increasing the tunneling, $\textrm{t}$, delocalizes the electrons in the central TQD molecule which decreases the contribution of the spin polarized configuration, Eq.~\ref{eq:dcfu}. But in this scenario even a small value of tunneling t' between the central and auxiliary dots can drive the GS in the maximally spin polarized subspace and so the area comprising of GS in total spin S=$\frac{3}{2}$ increases.\\

\indent It is also observed that the maximum value of overlap probability lies along the border line dividing the regions where the GS is in the subspace of total spin S=$\frac{1}{2}$ (zero probability) and that of total spin S=$\frac{3}{2}$ (finite probability).\\

\begin{figure}[htbp]
\centering
\vspace*{0.6cm}
\includegraphics[height=7.0cm,width=8.0cm]{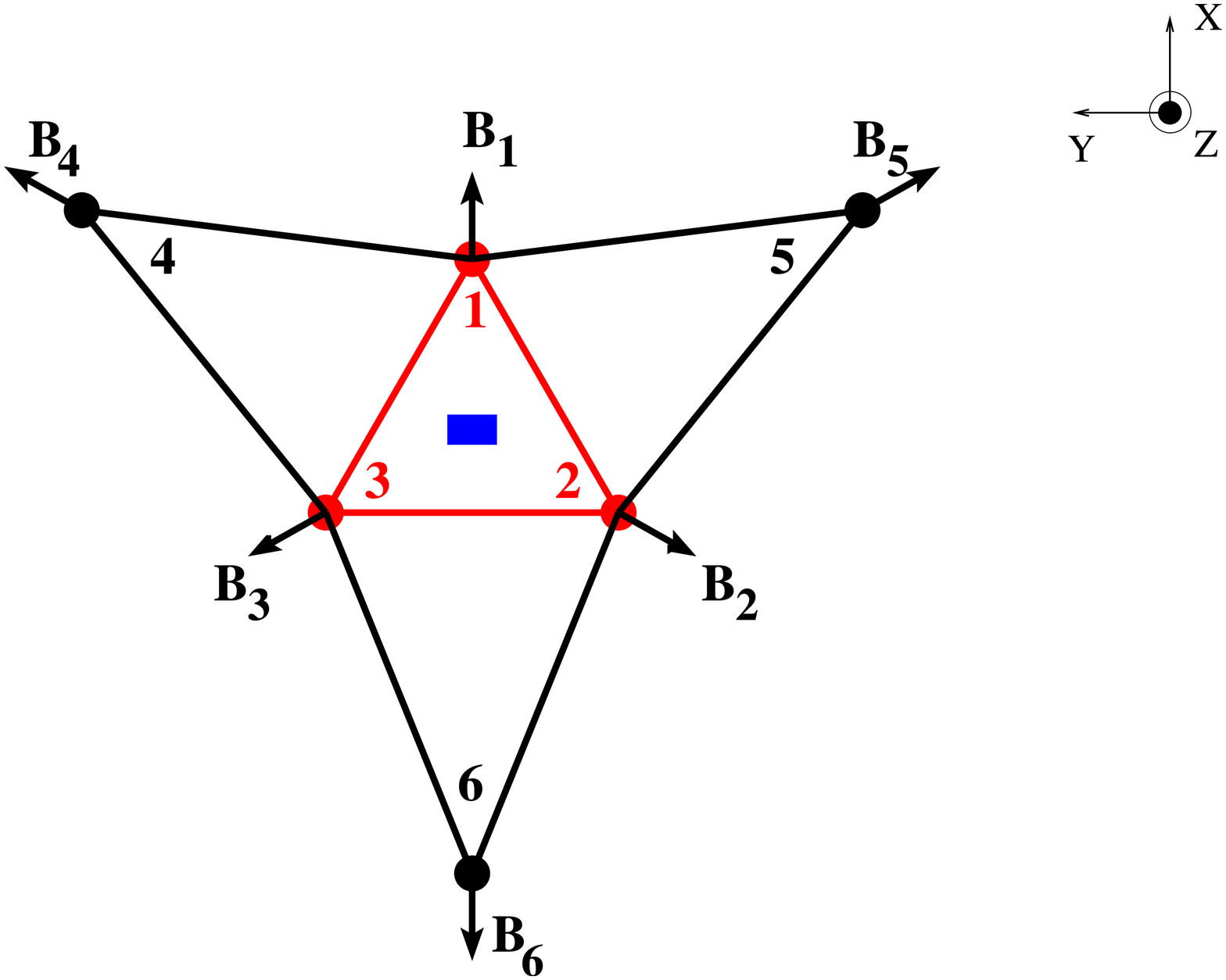}
\vspace*{0.4cm}
\caption{\label{fig:sqdmwbf} (Color online) Schematic depicting top bar magnets (rectangular box shown in blue) with like poles facing each other are placed at the center of the system which generates a radial in- plane magnetic field.}
\end{figure}

\indent Choosing an appropriate set of parameters  can maximize the overlap probability and prepare the GS of the system with spin polarized  charge configuration given in Eq.~\ref{eq:dcfu}. We thus define a GHZ state for the SQDM system as,

\begin{equation}\label{eq:ghz}
|\mathcal{GHZ}^{\pm}> = \frac{1}{\sqrt{2}} [ |\textrm{US}> \pm |\textrm{DS}> ].
\end{equation}

\indent In order to form one of the states given in Eq.~\ref{eq:ghz} as a GS of the SQDM we apply a radial in- plane magnetic field  as shown in Fig~\ref{fig:sqdmwbf}. The Hamiltonian describing this interaction of the electron spins with external magnetic field is, $\mathcal{H}^{\textrm{B}} = \textrm{g}_{\textrm{e}}\mu_{\textrm{B}} \sum_{\textrm{i=1}}^{6} \textbf{B}_{\textrm{i}} \cdot \textbf{S}_{\textrm{i}}$ and is given as

\begin{align}\label{eq:radhamsqd}
\mathcal{H}^{\textrm{B}} \nonumber
& = \textrm{b} \bigg[ \sigma^{\textrm{x}}_{1} \bigg] + \textrm{b} \bigg[ -\frac{1}{2}\sigma^{\textrm{x}}_{2} -\frac{\sqrt{3}}{2}\sigma^{\textrm{y}}_{2} \bigg] + \textrm{b} \bigg[ -\frac{1}{2}\sigma^{\textrm{x}}_{3} +\frac{\sqrt{3}}{2} \sigma^{\textrm{y}}_{3} \bigg] \nonumber \\
& + \textrm{b} \bigg[ \frac{1}{2}\sigma^{\textrm{x}}_{4} +\frac{\sqrt{3}}{2}\sigma^{\textrm{y}}_{4} \bigg] + \textrm{b} \bigg[ \frac{1}{2}\sigma^{\textrm{x}}_{5} -\frac{\sqrt{3}}{2} \sigma^{\textrm{y}}_{5} \bigg] + \textrm{b} \bigg[ -\sigma^{\textrm{x}}_{6} \bigg].
\end{align}

\indent The spin operators ($\sigma^{\textrm{x}}$, $\sigma^{\textrm{y}}$) are written in second quantized form as

\begin{equation}
\sigma^{\textrm{x}}_{\textrm{j}} = \frac{1}{2} [\textrm{c}_{\textrm{j}\uparrow}^{\dagger} \textrm{c}_{\textrm{j}\downarrow} + \textrm{c}_{\textrm{j}\downarrow}^{\dagger} \textrm{c}_{\textrm{j}\uparrow}]
\end{equation}

\indent and

\begin{equation}
\sigma^{\textrm{y}}_{\textrm{j}} = \frac{i}{2}  [\textrm{c}_{\textrm{j}\downarrow}^{\dagger} \textrm{c}_{\textrm{j}\uparrow} - \textrm{c}_{\textrm{j}\uparrow}^{\dagger} \textrm{c}_{\textrm{j}\downarrow}].
\end{equation}

\indent We write the complete Hamiltonian for SQDM and an applied external magnetic field as,

\begin{equation}\label{eq:fullhamsqd}
\mathcal{H} = \mathcal{H}^{\textrm{QDM}} + \mathcal{H}^{\textrm{B}}
\end{equation}

and the Hamiltonian matrix is constructed using the same basis as used to create $\mathcal{H}^{\textrm{SQD}}$. But since the magnetic field doesn't preserve the spin rotational symmetry the Hamiltonian matrix in Eq.~\ref{eq:fullhamsqd} is not in a block diagonal form. The dimension of the full Hilbert space is 220. We diagonalize the Hamiltonian matrix for three different values of $\epsilon$ and tuneling matrix element, $\textrm{t}$. These are chosen such that in the absence of an external magnetic field the area covered by the subspace of total spin S=$\frac{3}{2}$ in each view graph of Fig.~\ref{fig:phdia} is i) minimum [$\epsilon$=-0.10,t=0.04)], ii) intermediate [$\epsilon$=-0.05,t=0.05] and iii) maximum [$\epsilon$=0.00,t=0.06]. And for each value of $\epsilon$ and $\textrm{t}$, we diagonalize the Hamiltonian matrix for the parameters ($\frac{\textrm{V'}}{\textrm{V}}$,$\frac{\textrm{t'}}{\textrm{t}}$) where the overlap probability, Eq.~\ref{eq:ovpwtbf}, is maximum. \\

\indent At these points the low energy spectrum of extended Hubbard Hamiltonian resembles the energy spectrum of an isotropic Heisenberg Hamiltonian with four- fold degenerate ground and excited state as discussed in the previous section. For zero magnetic field the energy separation between these two degenerate state is $\frac{3\textrm{J}}{2}$. We use this to find the values of ferromagnetic exchange, $\textrm{J}$, in our Hubbard model for parameters (t,t',V,V',$\epsilon$) mentioned above with fixed U,V and we use it as our energy scale.\\

\begin{figure}[htbp]
\centering
\vspace*{0.6cm}
\includegraphics[height=6.0cm,width=8.0cm]{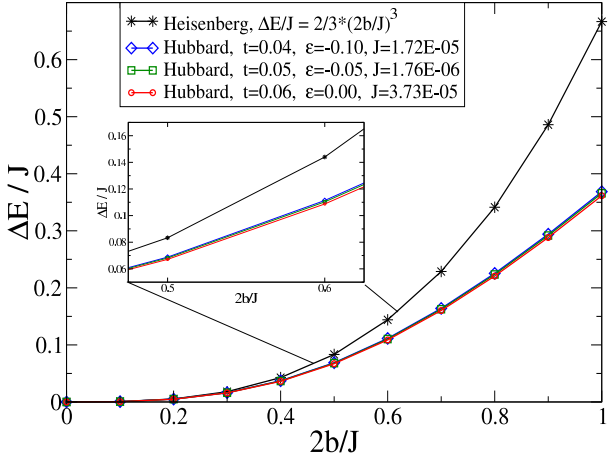}
\vspace*{0.4cm}
\caption{\label{fig:egwbf} (Color online) Energy splitting between the two GHZ states is shown as a function of $\frac{\textrm{2b}}{\textrm{J}}$ for three different values of $\epsilon$, t and J. The stars represent the results as obtained for three spin system, Heisenberg model.}
\end{figure}

\indent For each set of parameters, exact diagonalization of Eq.~\ref{eq:fullhamsqd} yields the two GHZ states as the ground and first excited states of our system. In Section~\ref{sec:tpsys} for Heisenberg system, we have seen that the energy separation between the two GHZ states for finite magnetic field is $\approx \frac{16\textrm{b}^{3}}{3\textrm{J}^{2}}$ where b is the effective strength of magnetic field. Apart from Heisenberg model, we calculate the energy splitting between the two GHZ states for our Hubbard model as a function of $\frac{\textrm{2b}}{\textrm{J}}$ for three different values of $\epsilon$, t and corresponding J. The result is shown in Fig.~\ref{fig:egwbf}. We observe that for smaller values of $\frac{\textrm{b}}{\textrm{J}}$, Heisenberg and Hubbard model behaves similarly but for larger values the agreement is worse. But it is also seen (inset of Fig.~\ref{fig:egwbf}) that as we keep on decreasing $\epsilon$ and $t$ the energy gap obtained from the Hubbard model starts approaching the Heisenberg model. This is expected since the value of maximum overlap probability is smallest for ($\epsilon$=0.00,t=0.06) and largest for ($\epsilon$=-0.10,t=0.04).\\

\begin{figure}[htbp]
\centering
\vspace*{0.6cm}
\includegraphics[height=6.0cm,width=8.0cm]{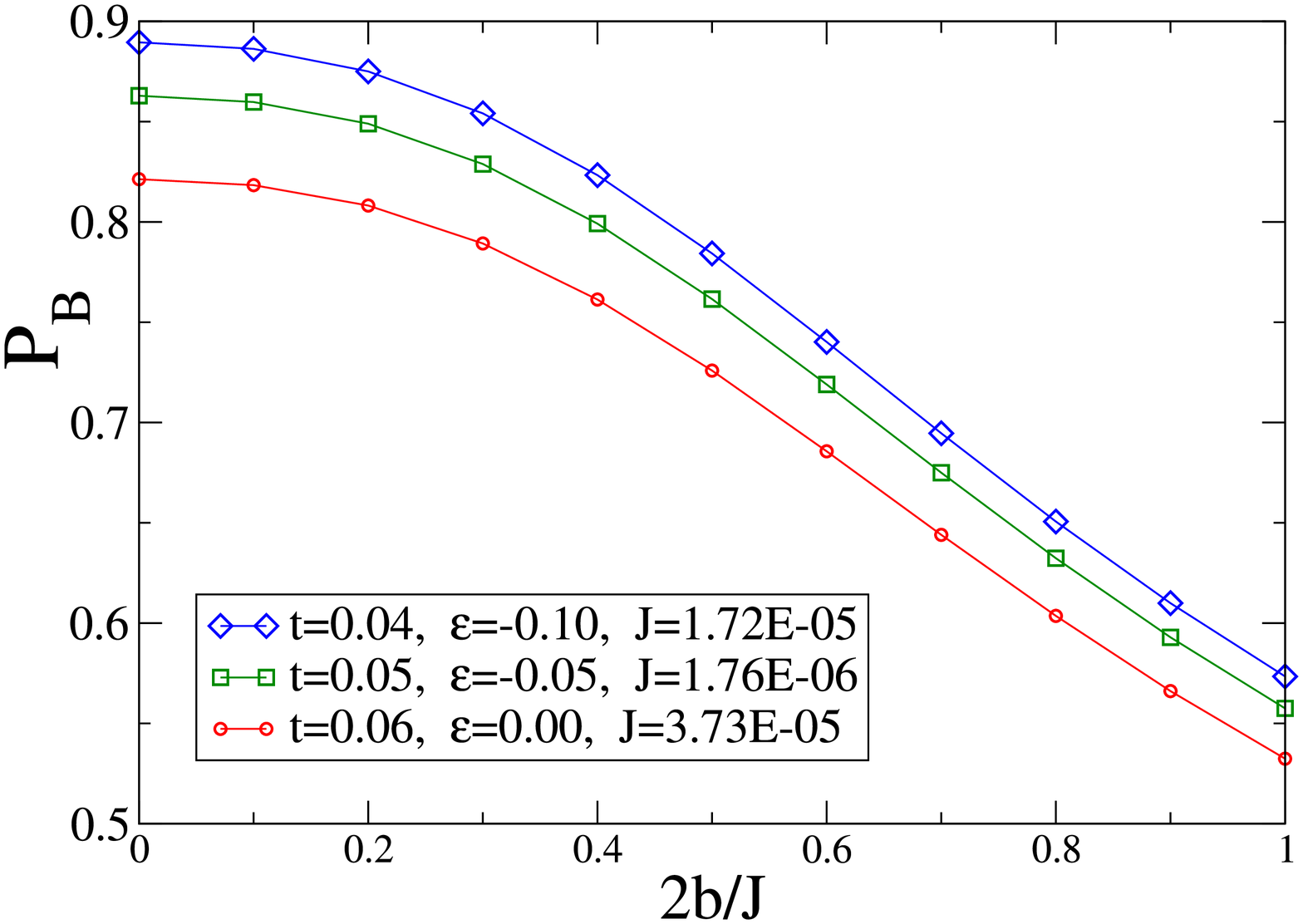}
\vspace*{0.4cm}
\caption{\label{fig:olpgswbf} (Color online) Maximum overlap probability of the ground state with $\mathcal{GHZ}^{+}$ is shown as a function of $\frac{\textrm{2b}}{\textrm{J}}$ for three different values of $\epsilon$, t and J.}
\end{figure}

\indent Using the same set of Hubbard parameters, we also evaluate the maximum overlap probability of the ground state with one of the GHZ state for the SQDM as given in Eq.~\ref{eq:ghz} for finite magnetic field and denote it as

\begin{equation}\label{eq:ovpwbf}
\textrm{P}_{\textrm{B}\neq 0} = |<\textrm{GS}|\mathcal{GHZ}^{+}>|^{2}
\end{equation}

\indent The results are shown in Fig.~\ref{fig:olpgswbf} as a function of $\epsilon$, $\textrm{t}$ and $\textrm{J}$. We find that as the strength of the effective magnetic field, b, approaches the value of J the GS of the Hubbard model deviates more and more from the GHZ state for SQDM which results in the decrease of the overlap probability.\\

\indent We now turn to the discussion of physical parameters and structures needed for the realization of GHZ generator. Typical parameters leading to an effective ferromagnetic coupling of three spins localized on a triangle, shown in Fig.~\ref{fig:phdia}, involve U=2, V=0.1, t=0.05, t'$\leq$t, V'$\leq$V and a bias $\epsilon$ of the order of tunneling, t. The values currently available for lateral gated quantum dots on GaAs are in the required parameter ranges of U$\sim$2meV, V$\sim$0.1meV, and t$\sim$0.05meV. By building quantum dot networks using individually gated self-assembled quantum dots on nanotemplates~\cite{dalacu} one can envisage reaching values of parameters for self-assembled quantum dots U$\sim$20meV, V$\sim$10meV, t$\sim$10meV which should lead exchange coupling reaching J$\sim$40meV, exceeding room temperature. \\

\indent Finally, we discuss the lifetime of a maximally entangled GHZ state. Since in the proposed scheme GHZ state is the ground state, it is expected to only suffer decoherence and no population decay. If the quantum dot molecule is realized in GaAs, nuclear spins are expected to be the major source of decoherence.  Since the problem of decoherence of the GHZ state is common with a single electron spin, recently developed coherent control of nuclear spins~\cite{folettibluhm} might be expected to be applicable in extending coherence of the GHZ state. This problem will be investigated in the future.\\

\section{\label{sec:sumconc} Summary $\&$ Conclusion}

In summary, we designed and analyzed theoretically a lateral quantum dot molecule combined with a micro-magnet generating a maximally entangled three particle GHZ groundstate. The quantum dot molecule consists of three quantum dots with one electron spin each forming a central equilateral triangle. The anti-ferromagnetic spin-spin interaction is changed to the ferromagnetic interaction by additional doubly occupied quantum dots, one dot near each side of a triangle. Exact diagonalization studies of the Hubbard model of the molecule determine the phase diagram in parameter space and a set of parameters is established for which the ground state of the molecule in a radial magnetic field is well approximated by a GHZ state. \\

\section{\label{sec:ack} Acknowledgment} 

\indent The authors thank QuantumWorks and Canadian Institute for Advanced Research for financial support. A.S. would like to thank C. -Y. Hsieh for fruitful discussions and his careful reading of the manuscript.


\begin{thebibliography}{99}
\bibitem{note1} In Fig.~\ref{fig:tmod} we assume a field direction pointing radially outward by considering two north poles facing each other. But a field direction pointing inwards by making the two south poles face each other can also be considered.
\end{thebibliography}
\end{document}